\documentclass[proceedings, preprint]{rmaa}

\usepackage{paralist}

\usepackage{psfrag,color}
\usepackage{hyperref}
 
\SetYear{2022}
\SetConfTitle{2nd Workshop on Astronomy Beyond the Common Senses}

\title{The universe in words: Astronomy for all through audio description within the outreach project Astroaccesible} 

\author{
  E. P\'erez-Montero,\altaffilmark{1} 
  C. Barn\'es-Casta{\~n}o,\altaffilmark{2}
  and E.J. Garc\'\i a G\'omez-Caro\altaffilmark{1}}

\altaffiltext{1}{Instituto de Astrof\'\i{}sica de Andaluc\'\i{}a - CSIC,
  Glorieta de la Astronom\'\i{}a s/n, 18008, Granada, Spain (epm@iaa.es, garcia@iaa.es).}

\altaffiltext{2}{Department of Translation and Interpreting, University of Granada, Granada, Spain  (cbc@ugr.es).}

\shortauthor{P\'erez-Montero, Barn\'es-Casta{\~n}o, \& Garc\'\i a G\'omez-Caro}
\shorttitle{Astronomy for all through audio description}

\listofauthors{E. P\' erez-Montero, C. Barn\' es-Casta{\~}no, \& E.J. Garc\'\i a G\'omez-Caro}
\indexauthor{P\' erez-Montero, E.}
\indexauthor{Barn\' es-Casta{\~}no, C.}
\indexauthor{Garc\'\i a G\'omez-Caro, E.J.}

\abstract{
	{\sc Astroaccesible} is an outreach project hosted by the Instituto de Astrof\'\i{}sica de Andaluc\'\i{}a (CSIC) and leaded by a blind astronomer. It is aimed at the teaching and popularisation of astronomy and astrophysics among all people regardless of their abilities using a wide range of inclusive resources. Audio description is one of the most extended inclusive resources for making films and artworks accessible for blind and partially sighted people. However, its potential for the description of astronomical images has not been widely exploited. In this contribution, we introduce  {\em The Universe in words}, a series of videos describing images of some of the most popular objects in the Messier catalogue. These audio descriptions do not only have a prominent inclusive aspect, but also imply a better and deeper understanding of the represented objects for everybody. The resulting videos can also be used as a complement to other resources for in-person activities, such as sonifications or tactile models of the same or similar astronomical objects.}

\resumen{
	{\sc Astroaccesible} es un proyecto de divulgaci\'on impulsado desde el Instituto de Astrof\'\i{}sica de Andaluc\'\i{}a (CSIC) y dirigido por un astr\'onomo ciego, cuyo objetivo es el de ense\~nar y divulgar la astronom\'\i{}a y la astrof\'\i{}sica entre todas las personas con independencia de sus capacidades mediante el uso de una amplia variedad de recursos inclusivos. La audiodescripci\'on se ha convertido en uno de los recursos inclusivos más extendidos, sobre todo con el fin de hacer accesibles pel\'\i{}culas y obras de arte para personas ciegas y con baja visión. Sin embargo, su potencial para describir im\'agenes astron\'omicas aún no ha sido desarrollado. En esta contribuci\'on presentamos {\em El Universo en palabras}, una serie de v\'\i{}deos en los que se describen im\'agenes de algunos de los objetos m\'as conocidos del cat\'alogo de Messier. Estas audiodescripciones no solo tienen un claro aspecto inclusivo, sino que tambi\'en conllevan una mejor y m\'as profunda comprensi\'on de los objetos representados para todo el p\'ublico. Los v\'\i{}deos resultantes tambi\'en pueden ser usados  como complemento a otros recursos,  en toda clase de actividades presenciales, como sonificaciones o maquetas de los mismos objetos astron\'omicos o similares.}

\addkeyword{Astronomy}
\addkeyword{science education}
\addkeyword{teaching material}
\addkeyword{teaching practice}

\begin{document}
\maketitle

\section{The Astroaccesible project}
\label{sec:sect1}

	The primary objective of the popularization and teaching of science for everyone
	can only be satisfactorily accomplished
	with the inclusion of all people regardless of their abilities, overcoming those barriers that prevent them
	totally accessing all the divulged contents.
	In the case of astronomy, blind and partially sighted people
	are excluded from having whole and satisfactory access
	to all its contents, given that visual media (e.g. pictures,
	videos, animations and graphics) play a leading role in the dissemination of 
	astronomical knowledge and research results.
	
	The project {\sc Astroaccesible} \footnote{\url{http://astroaccesible.csic.iaa.es}} \citep{pm17} was born with the aim of bringing astronomy closer to blind and partially sighted people by means of conferences,
	in-person classes, and the creation of adapted contents and
	resources to make the outreach of astronomy more accessible.
	The project has also a social dimension,
	as the principal investigator, Enrique P\'erez-Montero, is
	blind and is member of the Spanish National Organization for
	Blind People (ONCE). This helps to give visibility to
	scientists with disabilities conveying an idea of
	social inclusion in STEM areas \citep{pm19}.
	
	The main objective of in-person activities is to spark the interest for the
	study of astronomy for all challenging the idea that
	it can only be studied using the sense of sight \citep{galvez17}. This goal is achieved with
	different adapted resources and materials aiming at a complete description of several astronomical objects,
	their sizes, and relative distances using examples, helping
	blind and partially sighted people to acquire spatial notions using touch, complementing the verbal explanations and 
	also providing high-contrast images for those
	people with useful residual vision.
	This inclusive experience is enriched with more complex models, including hemispheres of the night sky in relief, or
	models of the Moon and other rocky planets of the Solar system \citep{ortiz19,ortiz20}.
	
	Despite these wide variety of options, the main resource for an accessible popularization of
	astronomy for blind and partially sighted people is a complete description
	of all elements, which can be either voiced or provided in paper format. This is the reason why
	we have focused on detailed descriptions of the concepts and
	contents presented in the conferences and in the articles available
	in the project webpage.
	This strategy is the most adequate from the point of view of
	blind and partially sighted people and it is the one that needs less
	additional resources.
	
	In addition, the situation caused by the COVID-19 pandemic in the last two years has forced many outreach events to move online, avoiding any
	direct interaction with touchable resources, and fostering the use of alternative strategies, such as oral descriptions, which can also be supported by other resources, such as sonifications
	(see also the contribution by Garc\'\i{}a-Benito et al. in this volume with the project {\sc Cosmonic}).
	In this contribution, we describe the project based on audio descriptions of astronomical images,
	called {\sc El Universo en palabras} ({\sc The Universe in words}), within the framework of the different resources designed for
	{\sc Astroaccesible} and that can be used both for in-person and virtual activities, with great level of acceptance from both
	sighted and blind people.

\section{Audio description as a resource}
\label{sec:sect2}

	Audio description is an access service in which images, be they dynamic (as in films) or static (as in paintings) are rendered into words. AD is regarded as an intersemiotic translation practice \citep{cnjh20}, as it involves the transfer of information that is conveyed by visual signs to linguistic ones. According to the Spanish AD standard, which includes guidelines on how to audio describe approved by the official Spanish standardization body, the main purpose of AD should be that of compensating for the lack of access to visual information in order that people with sight loss can comprehend the audio described source text (e. g. a film or a painting) in the most similar way as sighted people do. Although AD is mainly targeted at both blind and partially sighted people, these guidelines underscore that it has to be scripted with the information needs of congenitally blind people in mind. 
	
	AD practice and scholarship has so far focused on AD for films and audiovisual fiction, however it is gaining momentum in other areas, such as museums and live events \citep{barnes21}. Its application in science is still limited and it has only been explored from the production perspective, but its potential use in an astronomical context for both research and educational purposes is huge. 
	
	AD is also evolving in terms of the purposes and the audiences it might serve. For instance, it has been highlighted that AD can be successfully used in the foreign language classroom  and help people with emotion recognition difficulties. In the context of museum AD, Eardley et al. (2017) underline that “the standard visuocentric presentation of works of art assumes not only that people can see but also that they know how to use that vision to ‘look’ at a piece of art or a historical artifact to be able to extract some part of the rich cultural heritage associated with it” (17, 203).
However, research into how art expertise influences looking patterns suggests that they are driven by previous knowledge, with experts focusing more on formal features rather than on easily distinguishable elements \citep{koide15}. On the basis of these results, Eardley et al. (2017) propose that “guided looking may enhance the non-expert visitor’s experience”, thus directing visitors attention to those features that can be overlooked by the non-expert gaze. Drawing on studies from cognitive science, these authors also stress that this guided looking experience provided by the AD could also benefit sighted people as they would be exposed to a multisensory experience (at least visual and aural) which would help them not only to make sense of what they are seeing, but also to improved information processing and memorability. In a recent study, this claim has been empirically supported, as museum AD enhances the memorability of the museum experience also for sighted visitors as compared with traditional audio guides \citep{he21}. These encouraging results make it the more worthwhile to explore the potential of AD for being a truly inclusive resource benefiting blind and sighted people alike in all contexts where it can be applied, especially in those in which the lack of expert knowledge can negatively affect meaning-making processes, such as in astronomy.

\section{The Universe in words}
\label{sec:sect3}

	{\sc El Universo en palabras} is a series of audio described videos released in YouTube made in collaboration with the TRACCE research group of the University of Granada \footnote{\url{tracce.ugr.es}}.
	
	These audio descriptions were scripted by final-year Translation and Interpreting students as a group project for the subject Multimedia Translation during the courses 2019/2020 and 2020/2021. The main aim of this project is for students to put in practice all the competences they have acquired over the course of the subject and, at the same time, to apply this knowledge to a project which helps them delve into innovative translation modalities. Given the specialized nature of the source texts, images of astronomy objects, their project was conducted under the supervision of not only the AD trainer, but also of astronomers with an expertise in science dissemination strategies to ensure that the astronomical knowledge and terminology were accurately conveyed for the intended audience.
	 On the one hand, the scripts include all the relevant visual details of the audio described object, helping thus blind and partially sighted people to create a mental image of the object and serving as a guided looking experience for sighted people, as the depicted object and its elements are zoomed in and out in the video edition as a function of the AD script. On the other hand, all essential information to ensure understanding of astronomy concepts by non-experts is provided, which makes these guides a science dissemination resource both for blind and sighted people. 

	The program has focused so far on objects of the Messier catalogue, but it will very soon also cover other objects with or without tactile aids along with astronomical facilities with a view to expanding the scope of this project by including in-person activities.
	
	All the published videos are in Spanish, but it is foreseen to adapt them to other languages in the near future. An example in English using a synthetic voice is available in the presentation at the webpage of the workshop.
	
	The list of astronomical objects published so far are:
	\begin{itemize}
		\item \href{https://www.youtube.com/watch?v=gtuXMrF2OI4}{Messier~1. The Crab nebula}: a supernova remnant. This resource could be enriched with several available sonifications of the radio emission from the central pulsar and the emission of the surrounding gas.
		\item \href{https://www.youtube.com/watch?v=tCrUWLjLGjY}{Messier~13. The Hercules cluster}: a globular cluster. This resource can be supported using tactile aids.
		\item \href{https://www.youtube.com/watch?v=iGncLeZvxxg}{Messier~16. The Eagle nebula}: star-forming complex with very bright blue stars and surrounding gas clouds and pillars of dust and molecular gas. There are also sonifications of this object.
		\item \href{https://www.youtube.com/watch?v=Nw4uwRbiNLE}{Messier~51. The Whirlpool Galaxy}: an example of a face-on spiral galaxy with multiple star-forming knots. For this object there are sonifications and images in relief.
		\item \href{https://www.youtube.com/watch?v=IWnsSogrV00}{Messier~87. The Virgo A galaxy}: a giant elliptical galaxy with a radio jet emitting from the central supermassive black hole. This can be completed with available touchable models and sonifications.
	\end{itemize}

\acknowledgments{
We especially thank all the students who have participated in the elaboration of the scripts of the audio descriptions of {\em El Universo en palabras}: A. Le\'on, A. P\'erez Dom\'\i{}nguez, A. Gil L\'opez, A. Benito Aparicio, A. Herm\'an Carvajal, C. Salmer\'on Borja, C. Ghidini, E. Alonso Ochoa, E. Lacanna, L. Pastor Iradier, M. Burguillos, M. G\'omez Regalado and S. Guti\'errez Gull\'on.
We also thank the project {\em Estallidos de formaci\'on estelar en galaxias} AYA2016-79724-C4 for its financial support for the narration of the videos.
Participation of CBC in this project is supported by the Spanish Ministry of Science, Innovation and Universities grant FPU17/0490}


\begin{thebibliography}
		\bibitem[Barn\'es-Casta{\~n}o et al. (2021)]{barnes21} Barn\'es-Casta{\~n}o, C. et al. 2021, New Voices in Translation Studies, 25, 1-26.
\bibitem[Chica-N\'u{\~n}ez \& Jim\'enez-Hurtado (2020)]{cnjh20} Chica-N\'u{\~n}ez, A. \& Jim\'enez-Hurtado, 2020, Journal of Audiovisual Translation, 3(2), 264-285.
\bibitem[Eardley et al. (2017)]{e17} Eardley, A. et al. 2017, Inclusion, disability and culture, 195-207, Dordrecht: Springer
		\bibitem[Galvez et al.(2017)]{galvez17} Galvez, A., Ballesteros, F., Garc{\'\i}a-Frank, A., et al.\ 2017, European Planetary Science Congress
\bibitem[Hutchinson \& Eardley (2021)]{he21} Hutchinson, R. \& Eardley, A. 2021, Museum Management and Curatorship, 36(4), 427-446. https://doi.org/10.1080/09647775.2021.1891563
		\bibitem[Koide et al. (2015)]{koide15} Koide et al. 2015, PloS one, 10(2), e0117696
		\bibitem[Ortiz-Gil et al.(2019)]{ortiz19} Ortiz-Gil, A., Burguet-Castell, J., P{\'e}rez-Montero, E., 
		et al.\ 2019, EPSC-DPS Joint Meeting 2019
		\bibitem[Ortiz Gil(2020)]{ortiz20} Ortiz Gil, A.\ 2020, XIV.0 Scientific Meeting (virtual) of the Spanish Astronomical Society, 265
		\bibitem[P{\'e}rez-Montero(2019)]{pm19} P{\'e}rez-Montero, E.\ 2019, Nature Astronomy, 3, 114. doi:10.1038/s41550-019-0693-3
		\bibitem[P{\'e}rez-Montero et al.(2017)]{pm17} P{\'e}rez-Montero, E.J., Garc{\'\i}a G{\'o}mez-Caro, E., S{\'a}nchez Molina, Y., et al.\ 2017, Highlights on Spanish Astrophysics IX, 742
		
			\end{thebibliography}
\end{document}